# Macroscopic dielectric response of the metallic particles embedded in host dielectric medium

L.G. Grechko [a,*], K.W. Whites [b], V.N. Pustovit [a], V.S. Lysenko [c]

[a] *Institute of Surface Chemistry, NAS of Ukraine, prospekt Nauki 31, 252022 Kiev, Ukraine*
[b] *Department of Electrical Engineering, University of Kentucky, 453 Anderson Hall, Lexington, KY, USA*
[c] *Institute of Semiconductor Physics, NAS of Ukraine, prospekt Nauki 29, 252022 Kiev, Ukraine*

## Abstract

A theoretical approach is proposed to calculate an effective dielectric constant of a matrix disperse system (MDS) of metallic particles (spheres) randomly distributed and embedded in a uniform dielectric medium. Deviations from the well-known Maxwell–Garnett formula have been observed. The effective dielectric constant for different volume fractions of the embedded particles is considered as well as the dependence on pair interaction between particles, and the relation between sizes of each pair of particles. The problem is solved in the electrostatic approximation. © 2000 Elsevier Science Ltd. All rights reserved.

## 1. Introduction

The problem of the definition of an effective dielectric constant in a composite medium has been the subject of much interest in this century. A number of different solutions [1–3] have been proposed, but unfortunately, a theory which is able to describe and satisfy the recent experimental data [2] still does not exist until now. In this work, we are concerned with the theoretical description of the optical properties of disordered clusters of particles (spheres) randomly distributed in a host matrix with a dielectric constant $\varepsilon_0$. The effective dielectric constant $\tilde{\varepsilon}$ of such a system is often given by the Maxwell–Garnett formula at low concentrations of inclusions (metallic particles). A theoretical method, for the calculation of the effective dielectric constant of a (MDS) taking into account the pair interaction between inclusions of different radii, is proposed.

## 2. Dielectric constant of a matrix disperse system

Analogous to the method proposed in Refs. [4,5], we have developed an equation for the group expansion of the effective permittivity, taking into account the pair interaction between two particles [6–8]:

$$\frac{4\pi}{3} \frac{\tilde{\varepsilon}+2\varepsilon_0}{\tilde{\varepsilon}-\varepsilon_0} = \frac{1}{\sum_a n_a \alpha_a} - \frac{\frac{4\pi}{3}\sum_{a,b} n_a n_b}{(\sum_a n_a \alpha_a)^2} \\ \times \int_0^\infty R_{ab}^2 \Phi(\vec{R}_{ab}) \mathrm{d}\vec{R}_{ab} \left[ \beta_{ab}^{\mathrm{II}}(\vec{R}_{ab}) + 2\beta_{ab}^\perp(\vec{R}_{ab}) \right], \quad (1)$$

where $R_{ab}=|\vec{R}_a - \vec{R}_b|$, $\vec{R}_a$ and $\vec{R}_b$ are the origins of the spheres $a$ and $b$, respectively; $n_a = N_a/V$; $N_a, N_b, N_c, \ldots$ define the number of particles of each kind; and $V$ is the volume of system. The sum in Eq. (1) is taken for all kinds of particles. The function $\Phi(\vec{R}_{ab})$ is a two-particle distribution function which was further taken in the form

$$\Phi(\vec{R}_{ab}) = \begin{cases} 1 & R_{ab} > r_a + r_b, \\ 0 & R_{ab} \leqslant r_a + r_b, \end{cases} \qquad \alpha_a = \frac{\varepsilon_a - \varepsilon_0}{\varepsilon_a + 2\varepsilon_0} r_a^3$$

is the usual dipole polarizability of a particle of a given kind $a$. In Eq. (1), $\beta_{ab}^{\mathrm{II}}(\vec{R}_{ab})$ and $\beta_{ab}^\perp(\vec{R}_{ab})$ define the longitudinal and transverse parts of the tensor polarizability $\beta_{ab}(\vec{R}_{ab})$ of particle $a$ in the presence of particle $b$ in the external field $\vec{E}\sim \vec{E}_0 \mathrm{e}^{-i\omega t}$. On taking into account pairs of dipole–dipole interactions between particles, we get

---

* Corresponding author.

$$\beta_{ab}^{\mathrm{II}} = X_{10}^{(a)}(\vec{R}_{ab}) - \alpha_a - 2\frac{\alpha_a \alpha_b}{R_{ab}^3},$$
$$\beta_{ab}^{\perp} = X_{11}^{(a)}(\vec{R}_{ab}) - \alpha_a + \frac{\alpha_a \alpha_b}{R_{ab}^3}. \quad (2)$$

The coefficients $X_{10}^{(a)}(\vec{R}_{ab})$ and $X_{11}^{(a)}(\vec{R}_{ab})$ can be obtained from the solution to the electrostatic response for spheres $a$ and $b$ in the external field $\vec{E}_0$ (see Appendix A).

Consider a particular case where the number of particles of each kind $a$ and $b$ is the same, i.e. $n_a = n_b = n_0$ and $B = B_a = B_b$. We define a ratio of two sphere radii as $\Delta_{ab} = r_b/r_a$. From the general Eq. (1), for the case of the problem of two kinds of particles with radii $r_a = R$, $r_b = r$ ($\Delta = r/R < 1$), we have

$$\tilde{\varepsilon} = \varepsilon_0 \left\langle 1 + \frac{3f_0(1+\Delta^3)}{\frac{1}{B} - f_0(1+\Delta^3) - \frac{2}{3}f_0 D} \right\rangle, \quad (3)$$

where

$$D = \frac{1+\Delta^6}{1+\Delta^3} \ln \frac{8+B}{8-2B} + \frac{\Delta^3}{2(1+\Delta^3)}$$
$$\times \left[ \left( \Delta^{3/4} + \frac{1}{\Delta^{3/4}} \right)^2 \ln \frac{(1+\Delta)^3 + B\Delta^{3/2}}{(1+\Delta)^3 - 2B\Delta^{3/2}} \right.$$
$$\left. - \left( \Delta^{3/4} - \frac{1}{\Delta^{3/4}} \right)^2 \ln \frac{(1+\Delta)^3 - B\Delta^{3/2}}{(1+\Delta)^3 + 2B\Delta^{3/2}} \right] \quad (4)$$

and $f_0 = (4\pi/3)R^3 n_0$ is the volume fraction of particles of kind $a$, and $n_0$, its concentration. Eq. (3) at $\Delta = 1$ gives the well-known result [5,6]

$$\frac{\tilde{\varepsilon} + 2\varepsilon_0}{\tilde{\varepsilon} - \varepsilon_0} = \frac{1 - \frac{2}{3}f(\varepsilon - \varepsilon_0/\varepsilon + 2\varepsilon_0) \ln(3\varepsilon + 5\varepsilon_0/2\varepsilon + 6\varepsilon_0)}{f(\varepsilon - \varepsilon_0/\varepsilon + 2\varepsilon_0)}. \quad (5)$$

## 3. Discussion

We have presented a theoretical approach to consider the optical properties of an MDS. A method is proposed for the calculation of the effective dielectric constant of MDS based on an expanded formulation of the well-known Maxwell–Garnett law taking into account the pair interaction between particles. This method was applied to describe interactions between two spherical metallic particles of different radii. The obtained results allow us to predict the absorption spectral behavior of MDS in the neighboring infrared and visible regions of the spectrum in the region of the Maxwell–Garnett formula application.

The method presented in this article for the calculation of $\tilde{\varepsilon}$ has some restrictions. It is necessary that the volume fraction of inclusions in the matrix should not exceed the threshold of $f \leqslant 0.2$ [8,9]; otherwise, the calculated values from Eqs. (3) and (4) show considerable deviations from the experimental data. It should be noted that at $\Delta \to 0$, the main contribution to $\tilde{\varepsilon}$ gives the particles of large radii.

## Appendix A

Let us define $\vec{R}_a$ as a radius vector of the origin of the sphere $a$, and $\vec{R}$ as an arbitrary point in the medium. In the case of a uniform external field $\vec{E}_0$, the potential inside this sphere (which is regular at $1\vec{R} = \vec{R}_a$) has the form

$$\varphi_a^{\mathrm{in}} = -\vec{E}_0 \sum_{n'm'} A_{n'm'}^{(a)} |\vec{R} - \vec{R}_a|^{n'} Y_{n'm'}\left( \widehat{\vec{R} - \vec{R}_a} \right), \quad (A.1)$$

where $\widehat{\vec{R} - \vec{R}_a}$ is a unit vector along the direction $\vec{R} - \vec{R}_a$ and $Y_{n'm'}\left( \widehat{\vec{R} - \vec{R}_a} \right)$ is the spherical function. Potentials outside the sphere $a$ can be presented in the form

$$\varphi_a^{\mathrm{out}} = -\vec{E}_0 \sum_{n'm'} d_{n'm'} |\vec{R} - \vec{R}_a|^{n'} Y_{n'm'}\left( \widehat{\vec{R} - \vec{R}_a} \right)$$
$$- \vec{E}_0 \sum_{n'm'} B_{n'm'}^{(a)} |\vec{R} - \vec{R}_a|^{-n'-1} Y_{n'm'}\left( \widehat{\vec{R} - \vec{R}_a} \right)$$
$$- \vec{E}_0 \sum_{b \neq a} \sum_{n''m''} B_{n''m''}^{(b)} |\vec{R} - \vec{R}_b|^{-n''-1} Y_{n''m''}\left( \widehat{\vec{R} - \vec{R}_b} \right), \quad (A.2)$$

where the first term is the potential of the external field, the second term is the potential created by the particle $a$ in the point $\vec{R}$, and the third term is the potential from all the remaining particles in this point. In order to reduce the solution of this problem to one center, we shall use expressions for the transformation of the spherical functions from one center to another [10]:

$$\frac{Y_{n''m''}\left( \widehat{\vec{R} - \vec{R}_b} \right)}{|\vec{R} - \vec{R}_b|^{n''+1}} = \sum_{n'm'} Q_{n'm'}^{n''m''}(\vec{R}_b - \vec{R}_a)|\vec{R} - \vec{R}_a|^{n''} Y_{n'm'}\left( \widehat{\vec{R} - \vec{R}_a} \right),$$
$$Q_{n'm'}^{n''m''}(\vec{R}_b - \vec{R}_a) = (-1)^{n''+m''} \frac{Y_{n''+n',m'-m''}^{*}\left( \widehat{\vec{R}_b - \vec{R}_a} \right)}{|\vec{R}_b - \vec{R}_a|^{n''+n'+1}}$$
$$\times \left[ \frac{4\pi(2n''+1)(n''+n'+m''-m')!(n''+n'+m'-m'')!}{(2n'+1)(2n''+2n'+1)(n''+m'')!(n''-m'')!(n'-m')!(n'+m')!} \right]^{1/2} \quad (A.3)$$

where $|\vec{R} - \vec{R}_a| < |\vec{R}_b - \vec{R}_a|$.

Applying the solution of Eqs. (A.1) and (A.2) to the standard boundary conditions on the surface of inclusions and taking into account Eq. (A.3), we obtain a system of equations for the coefficients $B_{n'm'}^{(a)}$:

$$\frac{B^{(a)}_{n'm'}}{\alpha^{(a)}_{n'}} + \sum_{b \neq a}\sum_{n''m''} B^{(b)}_{n''m''} Q^{n''m''}_{n'm'}(\vec{R}_b - \vec{R}_a) = -d_{n'm'}. \quad (A.4)$$

In the case of two particles (*a* and *b*) with Eq. (A.4), we have

$$\frac{B^{(a)}_{n'm'}}{\alpha^{(b)}_{n'}} + \sum_{n''m''} B^{(b)}_{n''m''} (-1)^{m'+n''} \frac{Y_{n'+n'',m'-m''}(\widehat{\vec{R}_b - \vec{R}_a})}{R^{n'+n''+1}_{ab}} K^{n''m''}_{n'm'}$$
$$= -d_{n'm'},$$

$$\frac{B^{(a)}_{n'm'}}{\alpha^{(b)}_{n'}} + \sum_{n''m''} B^{(a)}_{n''m''} (-1)^{m'+n'} \frac{Y_{n'+n'',m'-m''}(\widehat{\vec{R}_b - \vec{R}_a})}{R^{n'+n''+1}_{ab}} K^{n''m''}_{n'm'}$$
$$= -d_{n'm'}, \quad (A.5)$$

Here, we define

$$K^{n''m''}_{n'm'} = \left[\frac{4\pi(2n''+1)(n'+n''+m'-m'')!(n'+n''+m''-m')!}{(2n'+1)(2n'+2n''+1)(n'+m')!(n'-m')!(n''+m'')!(n''-m'')!}\right]^{1/2} \quad (A.6)$$

and

$$\alpha^{(a)}_{n'} = \frac{n'\frac{\varepsilon_a}{\varepsilon_0} - 1}{n'\frac{\varepsilon_a}{\varepsilon_0} + n' + 1} r^{2n'+1}_a,$$

where $R_{ab} = |\vec{R}_b - \vec{R}_a|$, $r_a$ is the radius of particle *a*.

Further, we shall restrict ourselves only to the case of pair dipole–dipole interaction ($n' = n'' = 1$) and select the axis *OZ* along the vector $(\vec{R}_b - \vec{R}_a)$. Then, from Eqs. (A.2) and (A.5), we obtain expressions for the dipole moment of the particle *a*, given by

$$\vec{p}_a(a,b) \cdot \vec{m} = -E_0 \sum_{m'=-1} B^{(a)}_{1m'} Y_{1m'}(\widehat{\vec{R} - \vec{R}_a})$$
$$= \left[X^{(a)}_{10} n_z m_z + X^{(a)}_{11}(n_x m_x + n_y m_y)\right] E_0, \quad (A.7)$$

where

$$\vec{m} = \frac{\vec{R} - \vec{R}_a}{|\vec{R} - \vec{R}_a|}, \qquad \vec{n} = \frac{\vec{E}_0}{E_0},$$

$$X^{(a)}_{10} = \left[\frac{1}{\alpha^{(b)}_1} + \frac{2}{R^3_{ab}}\right]\left[\frac{1}{\alpha^{(a)}_1 \alpha^{(b)}_1} - \frac{4}{R^6_{ab}}\right]^{-1},$$
$$X^{(a)}_{11} = \left[\frac{1}{\alpha^{(b)}_1} - \frac{1}{R^3_{ab}}\right]\left[\frac{1}{\alpha^{(a)}_1 \alpha^{(b)}_1} - \frac{1}{R^6_{ab}}\right]^{-1}. \quad (A.8)$$

## References

[1] Bohren CF, Huffman DP. Absorption and scattering of light by small particles. New York: Wiley, 1983.
[2] Kreibig U, Vollmer M. Optical properties of metal clusters. Springer Series. In: Materials science, vol. 25. Berlin: Springer, 1995.
[3] Garland JG, Tanner PB, editors. Electrical transport and optical properties of inhomogeneous media. Ohio State University. New York: AIP, 1977.
[4] Felderhof BU, Ford GW, Cohen EGD. J Stat Phys 1982;28:649.
[5] Cichocki B, Felderhof BU. J Stat Phys 1988;53:499.
[6] Grechko LG, Levandovskii VG, Pinchuk AO. In: Optical storage. SPIE, vol. 3055. 1997. p. 111.
[7] Grechko LG, Pustovit VN. Proceedings of BIANISO-TROPICS'97. Glasgow, 1997. p. 227.
[8] Grechko LG, Blank AYu, Motrich VV, Pinchuk AO. Radiophys Radioastronomy 1997;2:19.
[9] Claro F. Phys Rev B 1984;B30:4989.
[10] Nozawa R. J Math Phys 1966;7:1841.